\documentclass[aps,prl,twocolumn,showpacs,groupedaddress]{revtex4}  
\usepackage{graphicx}  
\usepackage{dcolumn}   
\usepackage{bm}        
\usepackage{amssymb}   

\newcommand{\met}{\ensuremath{{\slash\kern-.7emE}_{T}}}
\newcommand{\vmet}{\ensuremath{\vec{\slash\kern-.7emE}_{T}}}

\newcommand{\mt}{\ensuremath{m_T}}

\newcommand{\wen}{\ensuremath{W \rightarrow e \nu}}

\newcommand{\pte}{\ensuremath{p_T^e}}
\newcommand{\ptnu}{\ensuremath{p_T^\nu}}

\newcounter{appendix}
\def\theappendix{\Alph{appendix}}
\def\appendix#1{
  \clearpage
  \addtocounter{appendix}{1}
  \addcontentsline{toc}{section}{\numberline{\theappendix}{#1}}
  \noindent{\bf\large Appendix \theappendix: #1}}

\newcounter {subsubsubsection}[subsubsection]

\def\pmcs{{\sc fastmc}}
\def\lsim{\mathrel{\rlap{\lower4pt\hbox{\hskip1pt$\sim$}}\raise1pt\hbox{$<$}}}

\begin{document}



\title{Measurement of the $W$ Boson Mass}
%
\author{V.M.~Abazov$^{37}$}
\author{B.~Abbott$^{75}$}
\author{M.~Abolins$^{65}$}
\author{B.S.~Acharya$^{30}$}
\author{M.~Adams$^{51}$}
\author{T.~Adams$^{49}$}
\author{E.~Aguilo$^{6}$}
\author{M.~Ahsan$^{59}$}
\author{G.D.~Alexeev$^{37}$}
\author{G.~Alkhazov$^{41}$}
% ...

\affiliation{\vspace{0.1 in}(The D\O\ Collaboration)\vspace{0.1 in}}
% ...
\affiliation{$^{6}$University of Alberta, Edmonton, Alberta, Canada;
                Simon Fraser University, Burnaby, British Columbia, Canada;
                York University, Toronto, Ontario, Canada and
                McGill University, Montreal, Quebec, Canada}
% ...
\affiliation{$^{30}$Tata Institute of Fundamental Research, Mumbai, India}
% ...
\affiliation{$^{37}$Joint Institute for Nuclear Research, Dubna, Russia}
% ...
\affiliation{$^{41}$Petersburg Nuclear Physics Institute,
                St. Petersburg, Russia}
% ...
\affiliation{$^{49}$Florida State University, Tallahassee, Florida 32306, USA}
% ...
\affiliation{$^{51}$University of Illinois at Chicago,
                Chicago, Illinois 60607, USA}
% ...
\affiliation{$^{59}$Kansas State University, Manhattan, Kansas 66506, USA}
% ...
\affiliation{$^{65}$Michigan State University,
                East Lansing, Michigan 48824, USA}
% ...
\affiliation{$^{75}$University of Oklahoma, Norman, Oklahoma 73019, USA}
% ...
\date{August 5, 2009}

\begin{abstract}
We present a measurement of the $W$ boson mass in $W\to e\nu$ decays using 
1~fb$^{-1}$ of data collected with the D0 detector during Run II of the
Fermilab Tevatron collider.
With a sample of 499830 $W \rightarrow e \nu$ candidate events, we measure 
$M_W = 80.401 \pm 0.043$~GeV.
This is the most precise measurement from a single experiment.
\end{abstract}

\maketitle

Knowledge of the $W$ boson mass ($M_W$) is currently a limiting factor in our ability
to tighten the constraints on the mass of the Higgs boson as determined from
internal consistency of the standard model (SM)~\cite{b:mwwa}. Improving the measurement of
$M_W$ is an important contribution to our understanding of the
electroweak (EW) interaction, and, potentially, of how the electroweak symmetry is
broken.  The current world-average measured value is $M_W =
80.399 \pm 0.025$~GeV~\cite{b:mwwa} from a combination of
measurements from the ALEPH~\cite{AlephW}, DELPHI~\cite{DelphiW}, L3~\cite{L3W},
OPAL~\cite{OpalW}, D0~\cite{D0W}, and CDF~\cite{CDFW,CDFNewW} collaborations.

In this Letter we present a measurement of $M_W$ using data collected from
2002 to 2006 with the D0 detector~\cite{d0det}, corresponding to a total
integrated luminosity of 1 fb$^{-1}$~\cite{lumi}.
We use the \wen\ decay mode because the D0 calorimeter is
well-suited for a precise measurement of electron energies, providing an energy
resolution of 3.6\% for electrons with an energy of $50$~GeV. The components of
the initial state total momentum and of the neutrino momentum along the beam
direction are unmeasurable, so $M_W$ is measured using three kinematic
variables measured in the plane perpendicular to the beam direction:
the transverse mass \mt, the electron transverse momentum $\pte$, and the
neutrino transverse momentum $\ptnu$.  The transverse mass is defined as
$
\mt = \sqrt{2\pte \ptnu(1-\cos\Delta\phi)}, \label{eqn:mt}
$
where $\Delta\phi$ is the opening angle between the electron and neutrino
momenta in the plane transverse to the beam.  The magnitude and
direction of $\ptnu$ are inferred from the event missing transverse energy ($\vmet$).
The $M_W$ measurement is made by comparing data spectra of $\mt$,
$\pte$, and $\met$ with probability density functions (templates) for these
spectra constructed
from Monte Carlo simulation with varying input $M_W$ values.

The D0 detector~\cite{d0det} contains tracking, calorimeter, and muon
systems.
Silicon microstrip tracking detectors (SMT) near the interaction point cover
pseudorapidity $| \eta | 
\lsim 3$ 
to provide
tracking and vertex information.  
%
The central fiber tracker surrounds the SMT, providing coverage to 
$| \eta | \approx 2$.
A 2 T solenoid surrounds these tracking detectors.
Three uranium, liquid-argon calorimeters measure particle energies.
The central calorimeter (CC) covers $| \eta | < 1.1$, and two end calorimeters
(EC) extend coverage to $| \eta | \approx 4$.  
The CC is segmented in depth
into eight layers.  The first four layers are used primarily to measure the
energy of photons and electrons and are collectively called the electromagnetic
(EM) calorimeter.  The remaining four layers, along with the first four, are
used to measure the energy of hadrons.  
Intercryostat detectors (ICD) provide added sampling in the region
$1.1 < |\eta | < 1.4$ where the CC and EC cryostat walls degrade the
calorimeter energy resolution.
%
%
%
%
A three level trigger system selects events for recording with a rate of 100 Hz.

Events are initially selected using a trigger requiring at least one EM cluster
found in the CC with transverse energy threshold varying from 20~GeV to 25~GeV
depending on run conditions.
Additionally, the position of the reconstructed
production point of a $W$ or $Z$ boson along the beam line is required
to be within $60$~cm of the center of the detector.

Candidate $W$ boson events are required to have one EM cluster reconstructed in
the CC, with $\pte>25$~GeV and $|\eta|<1.05$ where
$\eta$ is the pseudorapidity measured with respect to the center
of the detector.  The EM cluster must pass electron shower shape and energy isolation
requirements in the calorimeter, be within the central 80\% of the
electromagnetic section of each CC module, and have one track matching in
$(\eta,\phi)$ space, where the track has at least one SMT hit and $p_T>10$~GeV.
The central 80\% requirement is applied to the $\phi$ coordinate only and
excludes regions with slightly degraded energy resolution. The event must
satisfy $\met > 25$~GeV, $u_T < 15$~GeV, and $50 <
\mt < 200$~GeV.  Here $\met$ is the magnitude of the vector sum of the
transverse
energy of calorimeter cells above read out threshold, excluding those in the
coarse hadronic layer and in the inter--cryostat detector, and $u_T$ is the magnitude of the vector sum
of the transverse component of the energies measured in calorimeter cells
excluding those associated with the reconstructed electron.  This selection
yields 499,830 candidate $W\to e\nu$ events.  Throughout this Letter we use
``electron'' to imply either electron or positron.

We use $Z\to ee$ events for calibration. Candidate $Z$ boson events are
required to have two EM clusters satisfying the requirements above.  Both
electrons must have $\pte>25$~GeV.  One must be reconstructed in the CC and the
other in either the CC or EC ($1.5<|\eta|<2.5$).  The associated
tracks must be of opposite charge.  Events must also have $u_T < 15$~GeV and
$70~\mathrm{GeV}
\le m_{ee}
\le 110$~GeV, where $m_{ee}$ is the invariant mass of the dielectron
pair.  
Events with both electrons in the CC are used to determine the EM calibration.
There are 18,725 candidate $Z\to ee$ events in this category.

The backgrounds in the $W$ boson sample are $Z\to ee$ events in
which one electron escapes detection, multijet
events (MJ) in which a jet is misidentified as an electron with $\met$ arising
from misreconstruction, and $W\to \tau\nu \to e\nu\nu\nu$ events.  The background
from $Z$ boson events arises from electrons which traverse the gap between the
CC and EC.  The tracking efficiency in this region is high,
so this background is estimated by selecting data events passing the $W$
boson selection in which an additional track is pointing at the gap region.
The MJ background is determined using a sample obtained by removing
the track matching requirement for the electron candidates.  The probabilities for background and $W$
boson signal events in this sample to have a matching track are measured in control
samples.  The number of events in the sample without the track requirement and
the two probabilities are then used to determine the number of MJ background
events in the final $W$ boson sample.  The $W\to\tau\nu\to e\nu\nu\nu$ contribution is
determined from detailed simulation of the process using the D0 {\sc
geant}~\cite{b:geant}-based simulation.  The backgrounds expressed as a
fraction of the final sample are $(0.90\pm0.01)$\% from $Z\to ee$,
$(1.49\pm0.03)$\% from MJ, and $(1.60\pm0.02)$\% from $W\to\tau\nu\to e\nu\nu\nu$.

$W$ and $Z$ boson production and decay
kinematics are simulated using the {\sc resbos}~\cite{resbos} next-to-leading
order generator which includes non-perturbative effects at low boson $p_T$.  These
effects are parametrized
by three constants ($g_1$, $g_2$ and $g_3$) whose values are
taken from global fits to data~\cite{g2}.  The radiation of one or two photons
is performed using the {\sc photos}~\cite{photos} program.

Detector efficiencies and energy response and resolution for the electron and
hadronic energy are applied to the {\sc resbos+photos} events using a fast
parametric Monte Carlo simulation (\pmcs) developed for this analysis.  The
\pmcs\ 
parameters are determined using a combination of detailed simulation and
control data samples.  The primary control sample used for both the
electromagnetic and hadronic response tuning is $Z\to ee$ events.  $W$ boson events are
also used in a limited manner, as are events recorded in random beam crossings,
with or without requiring hits in the luminosity counters.

\begin{figure}[hbpt]
  \includegraphics[width=0.67\linewidth]{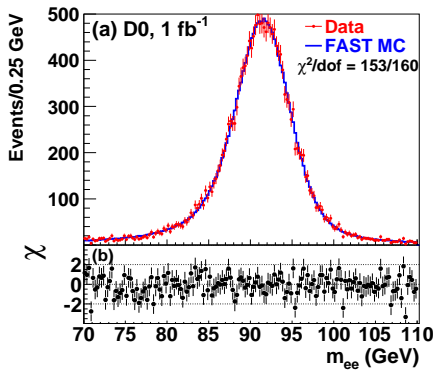}
  \caption{(a) The dielectron invariant mass distribution in $Z\to ee$ data and 
    from the fast simulation \pmcs\ and (b) the $\chi$ values where $\chi_i 
    = [N_i-\,($\pmcs$_i)]/\sigma_i$ for each point in the distribution, $N_i$ is
    the data yield in bin $i$ and $\sigma_i$ is the 
    statistical uncertainty in bin $i$.\label{f:zfinal}}
\end{figure}

Since the $Z$ boson mass and width are known with high precision from
measurements~\cite{ZLEP} at the CERN $e^+e^-$ collider (LEP), these values are used to calibrate the
electromagnetic calorimeter response assuming a form $E^{\text{meas}} =
\alpha\,E^{\text{true}} + \beta$ with $\alpha$ and $\beta$ constants determined
by calibration.  The $M_W$ measurement presented here is
effectively a measurement of the ratio of $W$ and $Z$ boson masses.
Figure~\ref{f:zfinal} shows a comparison of the $m_{ee}$ distributions for data
and \pmcs, as well as the $\chi$ distribution defined as the difference between
data and the \pmcs\ prediction divided by the statistical uncertainty on the difference.

The other major calibration is that of the hadronic energy in the event, which
includes energy recoiling against the boson.  The hadronic response (resolution)
is tuned using the mean (width) of the $\eta_{\text{imb}}$ distribution in $Z\to ee$
events in bins of $p_T^{ee}$.  Here $\eta_{\text{imb}}$ is defined as the sum
of the projections of the dielectron momentum 
$(\vec{p}_T^{ee})$ and $\vec{u}_T$ vectors in the transverse plane on the axis
bisecting the dielectron opening angle~\cite{ua2eta}.

To determine $M_W$, \pmcs\ template distributions for $\mt$, $\pte$, and
$\met$ are generated at a series of test $M_W$ values at intervals of
10~MeV with the backgrounds added to the simulated distributions.  A binned
likelihood between the data and each template is then computed.  The resulting
log likelihoods as a function of mass are fit to a parabola.  The minimum
point of the parabola defines the measured $M_W$ value.  The fits are performed
separately for each of the $\mt$, $\pte$, and $\met$ distributions,  and the
fit ranges were chosen to minimize the total 
expected uncertainty on $M_W$ for each distribution.

A test of the analysis procedure is performed using events produced by the
detailed {\sc geant} Monte Carlo simulation treated as collider data.  
The methods used for the data analysis
are applied to the simulated events, including the \pmcs\ tuning using the
simulated $Z\to ee$ events.  Each of the $M_W$ fit results using the $\mt$, $\pte$, and
$\met$ distributions agree with the input $M_W$ value within the
20~MeV total uncertainty of the test arising from Monte Carlo statistics.

\begin{figure*}[hbpt]
  \includegraphics[width=0.32\linewidth]{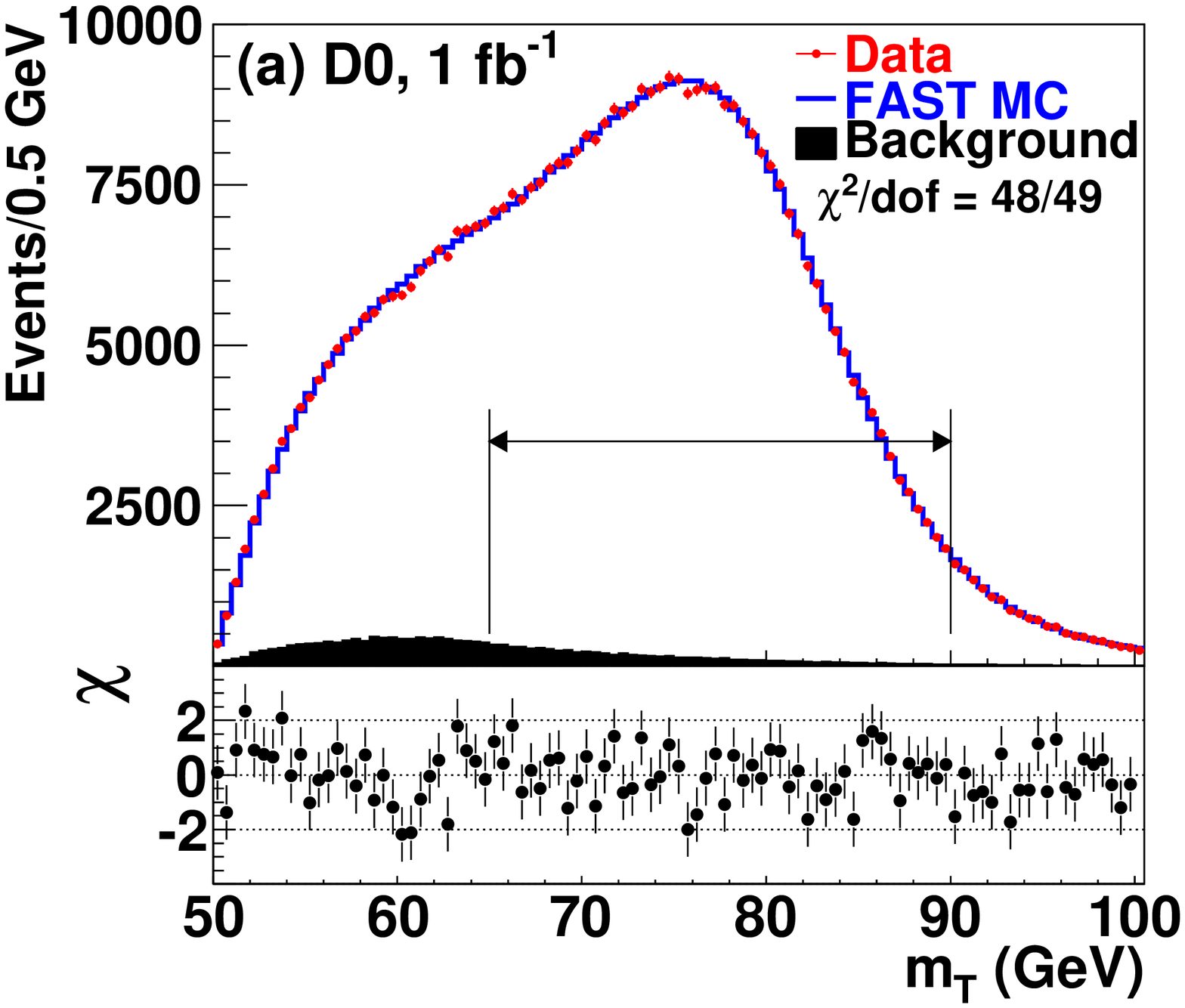}
  \includegraphics[width=0.32\linewidth]{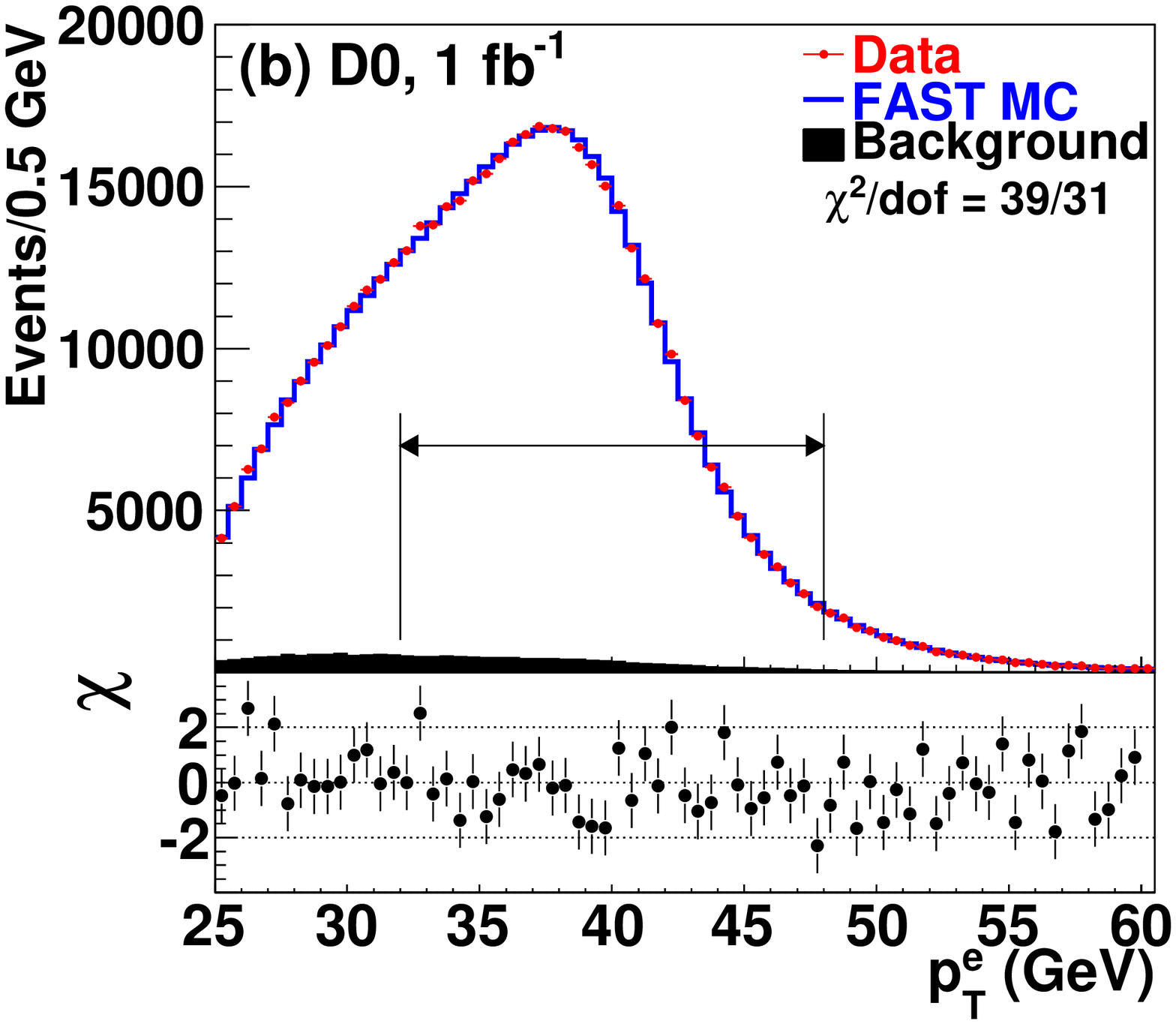}
  \includegraphics[width=0.32\linewidth]{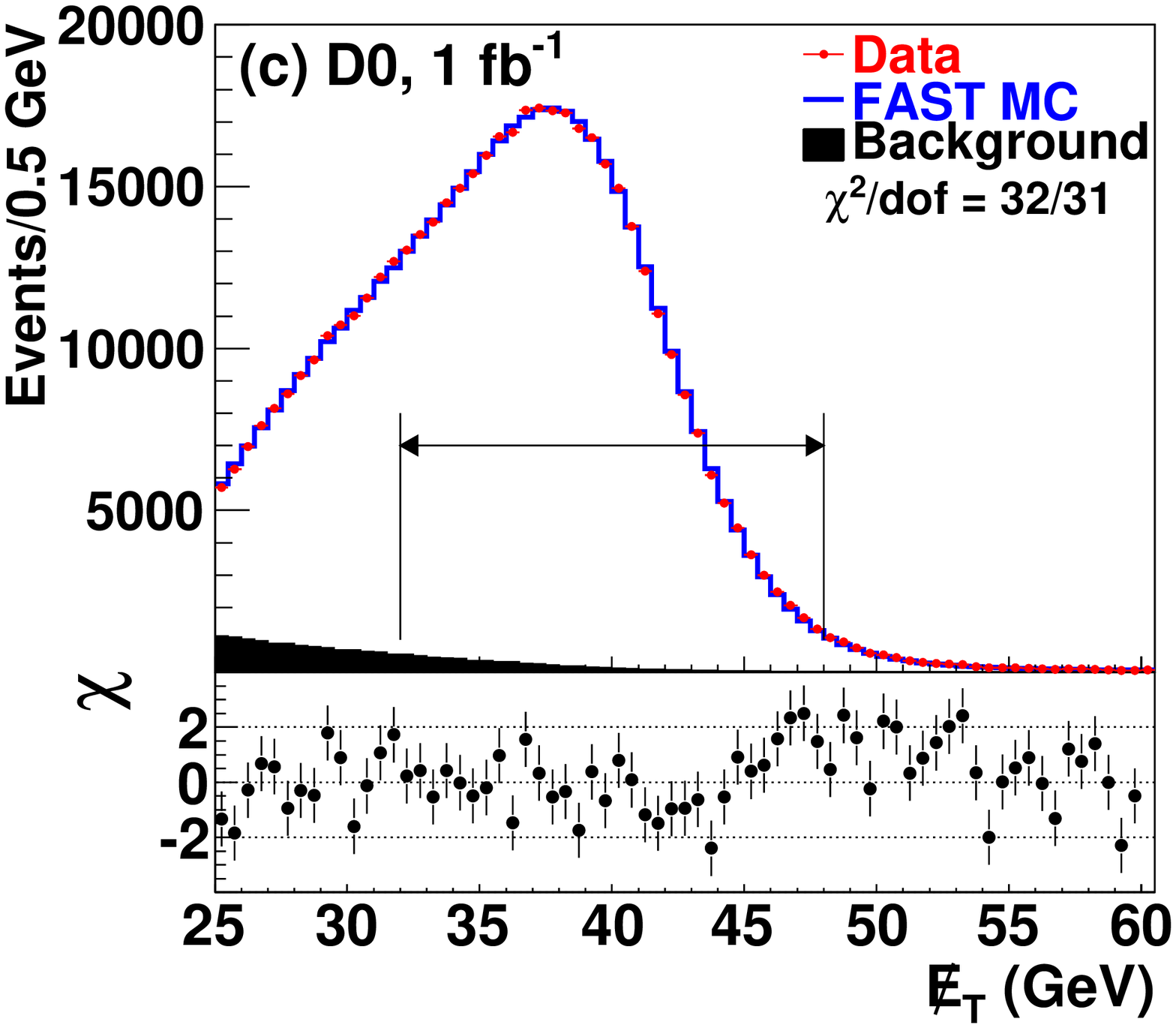}
  \caption{The (a) $\mt$, (b) $\pte$, and (c) $\met$ distributions for data and
    \pmcs\ simulation with backgrounds. The $\chi$ values are shown below each
    distribution where $\chi_i = [N_i-\,($\pmcs$_i)]/\sigma_i$ for each point in
    the distribution, $N_i$ is the data yield in bin $i$ and only the statistical uncertainty is used.  The fit
    ranges are indicated by the double-ended horizontal arrows.\label{f:mtfit}}
\end{figure*}

During the \pmcs\ tuning performed to describe the collider data, the $M_W$ values returned
from fits are blinded by the addition of an unknown constant offset.  The same
offset was used for $\mt$, $\pte$ and $\met$.  This
allowed the full tuning on the $W$ and $Z$ boson events and internal consistency
checks to be performed without knowledge of the final result.
Once the important data and \pmcs\ comparison plots have acceptable $\chi$
distributions, the results are unblinded.
The $Z$ boson mass value
from the post-tuning fit is $91.185\pm0.033\ \mathrm{(stat)}$~GeV, in
agreement with the world average of $91.188$~GeV used for the tuning.
The $M_W$ results from data after unblinding are given in
Table~\ref{t:answ}.  The $\mt$, $\pte$, and $\met$ distributions showing the data and \pmcs\ template
with background for the best fit $M_W$ are shown in Fig.~\ref{f:mtfit}.
\begin{table}[hbtp]
\begin{center}
  \caption{Results from the fits to data.  The uncertainty is only the
    statistical component.\label{t:answ}}
  \begin{tabular}{cccc}
     \hline\hline
     Variable & Fit Range (GeV) & $M_W$ (GeV)     & $\ \ \ \chi^2$/dof \\ \hline
      $\mt$   & $65<\mt<90$     & $\ \ \ 80.401\pm0.023\ \ \ $ &  48/49  \\
     $\pte$   & $32<\pte<48$    & $      80.400\pm0.027      $ &  39/31  \\ 
     $\met$   & $32<\met<48$    & $      80.402\pm0.023      $ &  32/31  \\
     \hline\hline
  \end{tabular}
\end{center}
\end{table}

The systematic uncertainties in the $M_W$ measurement arise from a
variety of sources, and can be categorized as those from experimental sources
and those from uncertainties in the production mechanism.  The systematic
uncertainties are summarized in Table~\ref{t:syst}.

The uncertainties on the electron energy calibration and the
hadronic recoil model are determined by simultaneously varying the parameters
determined in the tuning to $Z\to ee$ events by one statistical standard
deviation including correlation coefficients.  The electron energy resolution systematic
uncertainty is determined by varying resolution parameters determined in the
fit to the width of the observed $Z\to ee$ $m_{ee}$ distribution.  The shower
modeling systematic
uncertainties are determined by varying the amount of material representing the detector
in the detailed simulation within the uncertainties found by comparing the 
electron showers in the simulation
to those observed in data.  
No effect was seen when studying possible systematic bias for the energy loss
differences arising from the differing $E$ or $\eta$ distributions for the
electrons from $W$ and $Z$ boson decay.  The quoted systematic uncertainty is
due to the finite statistics of the event samples from the tuned detailed
simulation that are used to transport calibrations from the $Z$ to the $W$ sample.
The electron efficiency systematic is determined by
varying the efficiency by one standard deviation.
Table~\ref{t:syst} also shows the $M_W$ uncertainties arising from variation of the
background uncertainties indicated above.

Among the production uncertainties, the parton distribution function (PDF)
uncertainty is determined by generating $W$ boson events with the {\sc
pythia}~\cite{pythia} program using the CTEQ6.1M~\cite{cteq} PDF set.  The CTEQ
prescription~\cite{cteq} is used to determine a one standard deviation
uncertainty~\cite{CDFNewW} on $M_W$.  The QED uncertainty is
determined using {\sc wgrad}~\cite{wgrad} and {\sc zgrad}~\cite{zgrad}, varying
the photon-related parameters and assessing the variation in $M_W$ and by
comparisons between these and {\sc photos}.  
The boson $p_T$ uncertainty is determined by varying $g_2$ by its
quoted uncertainty~\cite{g2}.  Variation of $g_1$ and $g_3$ has negligible
impact.
\begin{table}
\begin{center}
  \caption{Systematic uncertainties of the $M_W$  measurement.\label{t:syst}}
  \begin{ruledtabular}
  \begin{tabular}{ lccc}
          & \multicolumn{3}{c}{$\Delta M_W$~(MeV)} \\
   Source                          &$\mt$ & $\pte$ &  $\met$\\
  \hline \hline
  Electron energy calibration       & 34 &  34 & 34 \\
  Electron resolution model         &  2 &   2 &  3 \\
  Electron shower modeling           &  4 &   6 &  7 \\
  Electron energy loss model        &  4 &   4 &  4 \\
  Hadronic recoil model             &  6 &  12 & 20 \\
  Electron efficiencies             &  5 &   6 &  5 \\
  Backgrounds                       &  2 &   5 &  4 \\ \hline
  Experimental Subtotal             & 35 &  37 & 41 \\ \hline
				    				     
  PDF                          &  10 &  11 & 11 \\
  QED                          &  7 &   7 &  9 \\
  Boson $p_T$                  &  2 &   5 &  2 \\ \hline
  Production Subtotal          & 12 &  14 & 14 \\ \hline

  Total                        &  37 & 40 & 43 \\
  \end{tabular}
  \end{ruledtabular}
\end{center}
\end{table}    

The quality of the simulation is indicated by the good $\chi^2$
values computed for the difference
between the data and \pmcs\ shown in the figures.  The data are also subdivided
into statistically independent categories based on instantaneous luminosity,
time, the total hadronic transverse energy in the event, the vector sum
of the hadronic energy, and electron pseudorapidity range.  The fit ranges are also varied.
The results are stable to within the measurement uncertainty for each of these
tests.

The results from the three methods have combined statistical and systematic
correlation coefficients of 0.83, 0.82, and 0.68 for $(\mt,\,\pte)$, $(\mt,\,\met)$, and
$(\pte,\,\met)$ respectively.  The correlation coefficients are determined using
ensembles of simulated events.  The results are combined~\cite{b:BLUE} including
these correlations to give the final result
\begin{eqnarray*}
  M_W & = & 80.401 \pm 0.021\ \mathrm{(stat)} \pm 0.038\ \mathrm{(syst)\ GeV}\\
      & = & 80.401 \pm 0.043\ \mathrm{GeV}.
\end{eqnarray*}
The dominant uncertainties arise from the available statistics of the $W\to
e\nu$ and $Z\to ee$ samples.  Thus, this measurement can still be expected to
improve as more data are analyzed.  The $M_W$ measurement
reported here agrees with the world average and the individual measurements and
is more precise than any other single measurement.  Its introduction in global
electroweak fits is expected to lower the upper bound on the SM
Higgs mass, although it is not expected to change the best fit
value~\cite{b:mwwa}.

%
We thank the staffs at Fermilab and collaborating institutions, 
and acknowledge support from the 
DOE and NSF (USA);
CEA and CNRS/IN2P3 (France);
FASI, Rosatom and RFBR (Russia);
CNPq, FAPERJ, FAPESP and FUNDUNESP (Brazil);
DAE and DST (India);
Colciencias (Colombia);
CONACyT (Mexico);
KRF and KOSEF (Korea);
CONICET and UBACyT (Argentina);
FOM (The Netherlands);
STFC and the Royal Society (United Kingdom);
MSMT and GACR (Czech Republic);
CRC Program, CFI, NSERC and WestGrid Project (Canada);
BMBF and DFG (Germany);
SFI (Ireland);
The Swedish Research Council (Sweden);
CAS and CNSF (China);
and the
Alexander von Humboldt Foundation (Germany).
%

\end{document}